\documentclass[conference,a4paper]{APSIPA2025}
\usepackage{amsmath}
\usepackage{graphicx}
\usepackage{multirow}
\usepackage{threeparttable}

\usepackage{geometry}
\usepackage{fancyhdr}

\fancypagestyle{firststyle}{
  \fancyhf{}
  \fancyhead[C]{2025 Asia Pacific Signal and Information Processing Association Annual Summit and Conference (APSIPA ASC)}
}

\usepackage{amsmath}
\usepackage{graphicx}
\usepackage{multirow}
\usepackage{threeparttable}
\usepackage{booktabs}
\usepackage{makecell}
\usepackage{adjustbox}
\usepackage{algorithm}
\usepackage{algpseudocode}
\usepackage{xcolor}  
\usepackage{placeins}
\usepackage{soul}
\usepackage{comment}
\usepackage{url}
\usepackage[hidelinks]{hyperref}
\begin{document}

\title{Speech Emotion Recognition via Entropy-Aware Score Selection}

\author{
\authorblockN{
ChenYi Chua\authorrefmark{1}, 
JunKai Wong\authorrefmark{1}, 
Chengxin Chen\authorrefmark{2},
Xiaoxiao Miao\authorrefmark{3}
}

\authorblockA{
\authorrefmark{1}
Singapore Institute of Technology, Singapore \\
E-mail: \{2302822, 2302765\}@sit.singaporetech.edu.sg
}

\authorblockA{
\authorrefmark{2}
Institute Of Acoustics, Chinese Academy Of Sciences, China
}

\authorblockA{
\authorrefmark{3}
Duke Kunshan University, China \\
E-mail: xiaoxiao.miao@dukekunshan.edu.cn
}
}
\maketitle
\thispagestyle{firststyle}
\pagestyle{fancy}

\begin{abstract}
In this paper, we propose a multimodal framework for speech emotion recognition that leverages entropy-aware score selection to combine speech and textual predictions. The proposed method integrates a primary pipeline that consists of an acoustic model based on wav2vec2.0 and a secondary pipeline that consists of a sentiment analysis model using RoBERTa-XLM, with transcriptions generated via Whisper-large-v3. We propose a late score fusion approach based on entropy and varentropy thresholds to overcome the confidence constraints of primary pipeline predictions. A sentiment mapping strategy translates three sentiment categories into four target emotion classes, enabling coherent integration of multimodal predictions. The results on the IEMOCAP and MSP-IMPROV datasets show that the proposed method offers a practical and reliable enhancement over traditional single-modality systems\footnote{Code can be found at \url{https://github.com/ExpiredTapWater/Emotion-Recognition}. This study is supported by Ministry of Education, Singapore, under its Academic Research Tier 1 (R-R13-A405-0005) and its SIT's Ignition grant (STEM) (R-IE3-A405-0005). Xiaoxiao Miao is the corresponding author and this work was conducted while she was at SIT.}.

\end{abstract}

\section{Introduction}
Speech Emotion Recognition (SER), which aims to recognise emotions directly from voice inputs as discrete emotion classes \cite{busso2008iemocap}, has become a crucial area of study in human-computer interaction, enhancing the emotional intelligence of virtual assistants, interactive robots, and mental health monitoring systems \cite{maji2023multimodal}. 
The rapid development of deep SER models, such as Convolutional Neural Networks (CNNs) \cite{bertero2017first}, Recurrent Neural Networks (RNNs) \cite{khalil2019speech}, and Transformer-based architectures \cite{baevski2020wav2vec, hsu2021hubert, chen2022wavlm}, has substantially improved recognition accuracy by capturing complex temporal and contextual patterns in speech. Despite these advances, SER remains challenging due to the subtlety and complexity of emotional expression, limited data availability, and ambiguous labeling, which often lead to misclassification \cite{poria2019emotion, Baltruvsaitis2018Multimodal}.

To address these issues, multimodal approaches that combine speech with textual or visual information have been explored to improve robustness \cite{poria2018multimodal, zadeh2017tensor}. Among these, integrating speech with text is particularly practical, as textual data can be obtained through automatic speech recognition (ASR) even when authentic transcripts are unavailable \cite{lu2020speech}. These transcripts are then processed using pretrained Transformer-based text models, such as BERT \cite{devlin2019bert} or RoBERTa \cite{liu2019roberta}, to extract rich textual features, while speech models are employed to extract acoustic features.

The key challenge lies in effectively fusing these two modalities. Fusion approaches can be broadly categorized into three types: Early fusion merges raw or low-level features from each modality but often struggles with alignment and dimensionality mismatches \cite{poria2018multimodal, katsaggelos2015audiovisual}. Intermediate fusion learns joint representations from both modalities, offering deeper integration but increasing training complexity \cite{zadeh2017tensor,chen2024modality}. Late fusion, or decision-level fusion, enhances flexibility by allowing each modality to operate independently before merging its outputs, reducing the impact of modality-specific errors \cite{georgescu2024exploring}. Techniques such as score averaging or rule-based merging further leverage the complementary strengths of different models while supporting independent updates to each component \cite{song2018decision}.

This work focuses on multimodal emotion recognition and proposes an entropy-aware late score selection strategy. 
A speech utterance is processed through two branches and obtains two scores. The primary speech branch utilizes a self-supervised learning model as a feature extractor, followed by a classifier that generates emotion predictions covering the full range of emotion classes, four classes in our experiments. The secondary textual branch processes the speech using an ASR model to obtain transcriptions, which are then fed into pretrained sentiment models applied off-the-shelf, without fine-tuning on emotion datasets to produce scores for three sentiment categories: Positive, Neutral, and Negative. 

The proposed entropy-aware score selection strategy guides the fusion of speech and sentiment scores. The first step is to evaluate the confidence of the speech score to determine whether intervention from the secondary model is necessary. This decision is based on whether the entropy and varentropy values of the speech score exceed delicately determined thresholds. If the speech score exhibits low confidence and low stability, the system refers to the secondary model for assistance. To address the mismatch between the four emotion classes and three sentiment categories, we introduce a sentiment mapping strategy: Positive and Neutral map directly to Happy and Neutral, while Negative is classified as Angry or Sad based on the primary model’s confidence. 

We evaluate these strategies on the IEMOCAP \cite{busso2008iemocap} and MSP-IMPROV \cite{busso2016msp} datasets to assess their impact on emotion classification performance. By incorporating the secondary model as a fallback mechanism for low-confidence predictions, we observed consistent improvements across both datasets, demonstrating enhanced classification precision and robustness, particularly in emotionally ambiguous cases. These results suggest that our late fusion framework offers a practical and reliable enhancement over a single-modality framework.

\begin{figure}[t]
  \centering
  \includegraphics[width=1\linewidth]{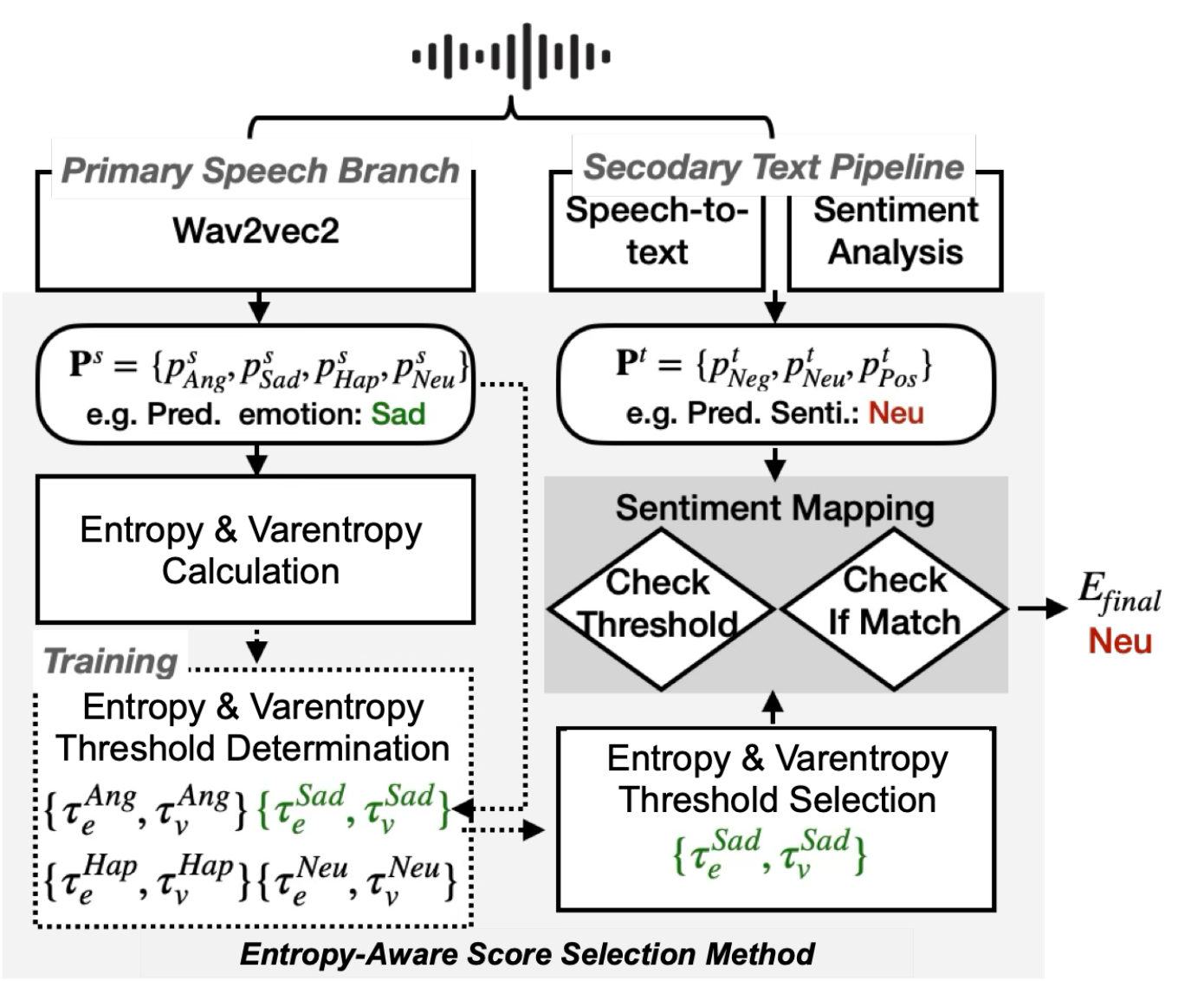}
  \caption{Overview of the proposed multimodal emotion recognition framework. The primary pipeline employs fine-tuned wav2vec2 to classify emotions directly from audio inputs into four emotion classes: Aug, Sad, Hap, and Neu. The secondary pipeline utilises Whisper-Large to transcribe audio into text, which is subsequently analysed for sentiment (Negative, Positive, Neutral) using RoBERTa XLM. Predictions from both pipelines are combined via the entropy-aware score selection strategy to produce the final predicted emotion.}
  \label{fig:speech_production}
  
\end{figure}

\section{Proposed Method}
This section details the proposed multimodal speech recognition approach, illustrated in Figure~\ref{fig:speech_production}, which comprises two independent pipelines processing the same audio input\footnote{Note that we could directly provide the authentic transcript. However, since obtaining the real transcript is labor-intensive in real-world applications, we assume that only speech is provided in this work, and the transcription is generated by a speech-to-text model.} to obtain predictions from two models, followed by an entropy-aware score selection strategy to boost performance.

\subsection{Primary Speech Modality}

The primary pipeline follows a conventional speech emotion recognition approach. A speech utterance is fed into a self-supervised learning-based wav2vec2 \cite{baevski2020wav2vec} as a feature extractor to obtain emotion-discriminative features,  which are then passed through a classifier consisting of two linear projection layers to generate emotion prediction. Specifically, we fine-tune wav2vec2 on the emotion datasets, with its parameters updated during the fine-tuning stage. The primary score can be represented as: $\mathbf{p}^s = \{ p^s_c \mid c \in \{\text{Ang, Sad, Hap, Neu}\}\}$, where $p^s_c$ represents the predicted probability for class $c$ among the four possible emotion classes in the primary branch. Since this branch is trained directly on speech data with labels, we consider it the primary/more reliable source for emotion prediction.

\subsection{Secondary Text Modality}
The secondary pipeline consists of a Speech-To-Text (S2T) model that generates transcribed text, which is then cleaned and preprocessed through several steps, including expanding contractions, removing punctuation, lemmatizing, lowercasing, and removing numbers, before being passed to a text-based sentiment analysis model \cite{camacho2017role}. This process does not involve any training on the specific speech datasets, using pretrained S2T and sentiment models applied off-the-shelf provides a second opinion on the expressed sentiment of the input speech.

\subsubsection{Whisper Series Speech-to-text Model}
Whisper is a robust multilingual transformer-based ASR model developed by OpenAI \cite{radford2023robust}. It is composed of an encoder-decoder architecture based on the Transformer framework. The encoder ingests log-Mel spectrogram features and consists of multiple stacked Transformer layers with multi-head self-attention, positional encodings, and layer normalization. The decoder generates the text output autoregressively, conditioned on both audio features and previously generated tokens.

\subsubsection{RoBERTa Series Sentiment Model}
We employ the RoBERTa series model for sentiment analysis \cite{barbieri2021xlm}. 
The model architecture consists of Transformer blocks, each with multi-head self-attention, feed-forward layers, residual connections, and layer normalization. The final hidden state is passed through a linear projection layer followed by softmax function to output probabilities across the three sentiment classes: Negative, Neutral, and Positive, denoted as $\mathbf{p}^t = \{ p^t_i \mid i\in \{\text{Negative, Neutral, Positive}\}\}$, where $p^t_i$ represents the predicted probability for class $i$ among the three possible sentiment classes in the secondary branch. 
This model is directly adopted from an open-source checkpoint trained on large-scale sentiment data. It is used without any adaptation or finetuning to provide a secondary opinion in score fusion.

\begin{algorithm}[t]
\footnotesize
\caption{\textnormal{$E_{\text{final}} = \text{Merge}(r, \tau_e^c, \tau_v^c, \tau_m^c, \mathcal{E}, \mathit{f_m}, \mathit{f_i})$}}
\label{alg:merge}
\begin{algorithmic}[1]
\Require
\Statex \textnormal{(i) A single sample $r$ with:}
\Statex \qquad \textnormal{– $r.\text{prediction}$: predicted emotion by the speech branch}
\Statex \qquad \textnormal{– $r.\text{sentiment}$: predicted sentiment by the text branch}
\Statex \qquad \textnormal{– $r.\mathbf{p}^s$: predicted class probabilities from the speech branch,}
\Statex \qquad \textnormal{\quad where $r.\mathbf{p}^s = \{p^s_{\text{Ang}}, p^s_{\text{Sad}}, p^s_{\text{Hap}}, p^s_{\text{Neu}}\}$}
\Statex \qquad \textnormal{– $r.\mathbf{p}^t$: predicted class probabilities from the text branch, }
\Statex \qquad \textnormal{\quad where $r.\mathbf{p}^t = \{p^t_{\text{Neg}}, p^t_{\text{Pos}}, p^t_{\text{Neu}}\}$ }
\Statex \qquad \textnormal{– $r.\mathcal{H}(\mathbf{p}^s)$: entropy of the speech prediction score}
\Statex \qquad \textnormal{– $r.\mathcal{V}(\mathbf{p}^s)$: varentropy of the speech prediction score}
\Statex \textnormal{(ii) \{Entropy, Valentropy, Mapping\} threshold sets $\{\tau_e^c, \tau_v^c\, \tau_m^c\}$ for each class, where $c \in \{\text{Ang, Sad, Hap, Neu}\}$}
\Statex \textnormal{(iii) A set $\mathcal{E}$ of disallowed emotion changes (Eg. {"AngSad", "NeuHap"})}
\Statex \textnormal{(iv) $\mathit{f_m}$: a string flag indicating the sentiment-to-emotion mapping strategy}
\Statex \textnormal{(iv) $\mathit{f_i}$: a string flag indicating to invert the final mapping}
\Ensure \textnormal{$E_{\text{final}}$: Final merged emotion}

\State $\tau_e, \tau_v, \tau_m \gets \tau_e^{r.\text{prediction}}, \tau_v^{r.\text{prediction}}, \tau_m^{r.\text{prediction}}$
\If{$r.\mathcal{H}(\mathbf{p}^s) \ge \tau_e$ \textbf{and} $r.\mathcal{V}(\mathbf{p}^s) \le \tau_v$}
    \If{$r.\text{sentiment} = \texttt{"neutral"}$}
        \State $emotion \gets \text{Neu}$
    \ElsIf{$r.\text{sentiment} = \texttt{"positive"}$}
        \State $emotion \gets \text{Hap}$
    \Else

        \If{$f_m = \texttt{"refer"}$} \Comment{Refer to primary model}
            \State $emotion \gets \text{Ang}$ \textbf{if} $r.p^s_{\text{Ang}} \ge r.p^s_{\text{Sad}}$ \textbf{else} $\text{Sad}$
        
        \ElsIf{$f_m = \texttt{"simple"}$} \Comment{Simple or flip mapping}
            \State $emotion \gets \text{Ang}$ \textbf{if} $ (r.{p}^t_{r.sentiment} \le \tau_m) \oplus \mathit{f_i} $ \textbf{else} $\text{Sad}$

        \EndIf
    \EndIf
    \If{$(r.\text{prediction},emotion) \in \mathcal{E}$}
        \State $emotion \gets r.\text{prediction}$ \Comment{Revert change}
    \EndIf
\Else
    \State $emotion \gets r.\text{prediction}$
\EndIf
\State $E_{\text{final}} \gets emotion$
\State \Return $E_{\text{final}}$
\end{algorithmic}
\end{algorithm}

\subsection{Entropy-Aware Score Selection Strategy}
\label{sec:entropy-aware}

After obtaining speech emotion $\mathbf{p}^s$ and sentiment $\mathbf{p}^t$ scores, an entropy-aware score selection strategy is introduced to determine when the primary model's predictions should be supplemented by those from the secondary model. 
The principle is that if the primary model’s prediction has higher certainty, it is retained, otherwise, the system relies on the secondary model. 

We utilize two metrics to quantify the certainty of the primary model’s prediction. The first metric is entropy $\mathcal{H}$ , directly measures the degree of uncertainty:
\begin{equation}
\mathcal{H} = \mathcal H(\mathbf{p}^s) = -\sum_{i=1}^{N} p^s_c \,\log(p^s_c).
\end{equation}
A lower $\mathcal{H}$ is preferred, indicating high prediction certainty.

The second metric is varentropy $\mathcal{V}$, which quantifies the dispersion of the class probabilities relative to the entropy:
\begin{equation}
\mathcal{V} = \mathcal{V}(\mathbf{p}^s) = \sum_{i=1}^{N} p^s_c  \left( \log(p^s_c ) + \mathcal{H}  \right)^2.
\end{equation}
Varentropy provides an additional measure of confidence by assessing how sharply peaked or flat the probability distribution is, thereby quantifying the stability of entropy $\mathcal{H}$. A higher varentropy $\mathcal{V}$ is preferred, indicating stable uncertainty estimates of the primary model, demonstrating robustness to minor input variations.

If a primary score $\mathbf{p}^s$ has high entropy $\mathcal{H}$ and low varentropy $\mathcal{V}$, measured against two thresholds $\tau_\mathcal{H}$ and $\tau_\mathcal{V}$, it is deemed unreliable. In such cases, we defer the final decision to the secondary sentiment model. This decision is implemented through the merge function described in Algorithm~\ref{alg:merge}, which maps the sentiment prediction into one of the target emotion classes. If the thresholds are not met, the original prediction from the primary model is used.

To derive the optimal thresholds \(\tau_e\) and \(\tau_v\), we perform a grid search over all possible pairs of entropy and varentropy values within empirically determined ranges on the training data.  
Note that the search for optimal thresholds is conducted separately for each emotion class \(c\), resulting in four sets of thresholds \(\{\tau_e^c, \tau_v^c\} \in [\{\tau_e^\text{Ang}, \tau_v^\text{Ang}\}, \{\tau_e^\text{Sad}, \tau_v^\text{Sad}\}, \{\tau_e^\text{Hap}, \tau_v^\text{Hap}\},  \{\tau_e^\text{Neu}, \tau_v^\text{Neu}\}]\).\footnote{We initially experimented with a single set of thresholds searched across the entire dataset, but the results were unsatisfactory. Upon analysis, we found that each class exhibits distinct entropy and varentropy distributions, which motivated the use of class-wise thresholds.}  
To do this, we group the training samples based on their emotion labels and determine the thresholds for each group independently. Taking one emotion class \(c\) as an example, suppose we have \(N\) training samples belonging to class \(c\). Let \(P_k(\mathcal{H})\) and \(P_k(\mathcal{V})\) denote the \(k\)-th percentiles of entropy and varentropy values for the \(N\) samples, respectively. The search ranges for the thresholds are defined as:
\begin{align}
\tau_e &\in \left[ P_{75}(\mathcal{H}) - \Delta, \quad P_{75}(\mathcal{H}) + \Delta \right], \\
\tau_v &\in \left[ P_{25}(\mathcal{V}) - \Delta, \quad P_{25}(\mathcal{V}) + \Delta \right].
\end{align}

These ranges are motivated by the observation that incorrect predictions tend to exhibit higher entropy and lower varentropy, with optimal threshold values typically located near the 75th percentile for entropy and the 25th percentile for varentropy. The search is conducted using a fixed step size of \(\Delta = \pm 10\) percentile points.
Each candidate threshold pair is given by:
\begin{align}
\tau_{\text{e}} = P_{75}(\mathcal{H}) - \Delta + k \delta, \quad
\tau_{\text{v}} = P_{25}(\mathcal{V}) - \Delta + l \delta,
\end{align}
for $k = 0, 1, \dots, K$ and $l = 0, 1, \dots, L$, forming a grid of candidate thresholds.
The optimal thresholds \((\tau_{\text{e}}^c, \tau_{\text{v}}^c)\) for each emotion class \(c\) are determined by maximizing a detection accuracy metric \(\mathcal{M}\), defined as:

\begin{align}
(\tau_{\text{e}}^c, \tau_{\text{v}}^c) = \arg\max_{(\tau_{\text{e}}, \tau_{\text{v}})} \mathcal{M}(\tau_{\text{e}}, \tau_{\text{v}}),
\end{align}
\noindent
where the objective metric \(\mathcal{M}\) is computed as:
\begin{align}
\label{eqn:metric}
\mathcal{M}(\tau_{\text{e}}, \tau_{\text{v}}) = \text{Accuracy} = \left( \frac{D}{T} \right) \big|_{\tau_{\text{e}}, \tau_{\text{v}}} \times 100,
\end{align}
where \( D \) denoting the number of misclassified samples successfully identified by the current thresholding rule, and \( T \) representing the total number of samples that satisfy both the entropy and varentropy threshold conditions.

\subsection{Sentiment Mapping Strategy}
This component addresses the challenge of mapping three-class sentiment outputs (Positive, Neutral, Negative) into the required four emotional classes (Happy, Neutral, Sad, Angry). While positive and neutral sentiments map straightforwardly to Happy and Neutral, respectively, negative sentiment needs further discrimination between Sad and Angry emotions. 
We propose two methods for this mapping, both of which are evaluated and the one yielding the higher accuracy is selected automatically.

\subsubsection{Refer to Primary Model Mapping}
In cases of negative sentiment, this method consults the primary model’s confidence scores between Angry and Sad 
emotions, assigning the sentiment accordingly based on the higher confidence score from the primary model. 

\subsubsection{Simple or Flip Mapping}
We establish a threshold (ranging between 0 and 1) $\tau_m^c$ for each emotion class $c$ where sentiment scores below the threshold map to Sad and those above to Angry. This mapping is flexible and can be reversed through a "flip" flag since both emotions can validly represent a negative sentiment. In a similar process to obtaining the optimal entropy and varentropy threshold values, a grid search on the training dataset is performed on a range of discreet values set at constant intervals using the same detection accuracy metric previously mentioned. 
\subsection{Revert Change Strategy}

Once all optimal threshold pairs \(\{\tau_e^c, \tau_v^c, \tau_m^c\}\) are determined, we adopt an additional strategy to enhance the prediction: we aim to construct an exclusion list \(\mathcal{E}\) using the training set, where all detrimental emotion changes are stored. Since these changes cause performance drops on the training set, during the inference stage, if such a change occurs, it will be skipped, and the primary prediction will be retained. 

Regarding the creation of \(\mathcal{E}\), for each training sample, if, for example, a sample's primary prediction is Angry, and after applying the entropy and valentropy thresholds, the prediction changes to Sad (indicated as AngSad), while the accuracy calculated using the same performance metric defined in Equation~\ref{eqn:metric} drops, this is considered a harmful change and will be added to \(\mathcal{E}\). 
Such changes should be avoided during the reference stage.

\begin{table}[ht]
\centering
\footnotesize
\caption{Emotion distribution in each session of the IEMOCAP and MSP-IMPROV datasets.}
\label{tab:datasets}
\begin{adjustbox}{max width=\linewidth}
\begin{tabular}{l|ccccc|cccccc}
\toprule
\multirow{2}{*}{\textbf{Emotion}} & \multicolumn{5}{c|}{\textbf{IEMOCAP Session No.}} & \multicolumn{6}{c}{\textbf{MSP-IMPROV Session No.}} \\
                                  & 1   & 2   & 3   & 4   & 5   & 1   & 2   & 3   & 4   & 5   & 6   \\
\midrule
Angry        & 229 & 137 & 240 & 327 & 170 & 54  & 54  & 73  & 52  & 119 & 108 \\
Happy        & 278 & 327 & 286 & 303 & 442 & 92  & 162 & 143 & 140 & 238 & 224 \\
Neutral      & 384 & 362 & 320 & 258 & 384 & 204 & 284 & 409 & 169 & 309 & 358 \\
Sad          & 194 & 197 & 305 & 143 & 245 & 76  & 78  & 73  & 76  & 109 & 215 \\
\midrule
\textbf{Total} & 1085 & 1023 & 1151 & 1031 & 1241 & 426 & 578 & 698 & 437 & 775 & 905 \\
\bottomrule
\end{tabular}
\end{adjustbox}
\end{table}

\section{Experiments}
In this section, we first select the speech-to-text model and sentiment model from various publicly available options, and then verify the effectiveness of the proposed entropy-aware score selection method on two commonly-used speech emotion recognition datasets.

\subsection{Datasets}
Two widely recognized datasets are used to verify the effectiveness of the proposed method. 
The first is IEMOCAP \cite{busso2008iemocap}, which comprises 5 sessions, each containing one male and one female actor performing scripted and improvised scenarios. We employ 10-fold cross-validation, where 4 sessions are used for training, and the utterances from the remaining session—containing two speakers—are used for validation and testing, respectively.

The second dataset, MSP-IMPROV \cite{busso2016msp}, consists of 6 sessions, each with one male and one female actor. We adopt 6-fold cross-validation, where each fold designates one complete session as the test set, while the remaining five sessions are split into training (80\%) and validation (20\%) subsets.

Table~\ref{tab:datasets} lists the number of utterances for each emotion in each session for both datasets.

\subsubsection{Settings}
For each cross-validation fold, we dynamically determine the optimal entropy, varentropy, and sentiment thresholds based on the training data of that fold, and then apply them to the corresponding test set. This approach ensures adaptive yet consistent prediction merging across all folds for both the IEMOCAP and MSP-IMPROV datasets.

\begin{table}[ht]
\caption{WER (\%) for different speech-to-text models (left) and F1 (\%) for Different Text-Based Sentiment Models using Ground Truth Transcripts (right).}

\label{tab:combined_performance}
\footnotesize
\centering
\begin{adjustbox}{max width=\linewidth}
\begin{tabular}{lc|lcccc}
\toprule
\textbf{S2T} & \textbf{WER} & \textbf{Sentiment} & \multicolumn{4}{c}{\textbf{F1 Score}} \\
\cmidrule(lr){4-7}
 &  && Neg. & Neu. & Pos. & Overall \\
\midrule
w2v-CTC\footnote{\url{https://huggingface.co/facebook/wav2vec2-base-960h}} & 34.26 & DistilBERT\footnote{\url{https://huggingface.co/lxyuan/distilbert-base-multilingual-cased}} & 11.62 & 2.75 & \textbf{76.39} & 17.79 \\
whisper-tiny\footnote{\url{https://huggingface.co/openai/whisper-tiny}} & 22.96 & RoBERTa\footnote{\url{https://huggingface.co/cardiffnlp/twitter-roberta-base-sentiment}}  & 48.09 & 46.60 & 43.66 & \textbf{46.12} \\
whisper-large\footnote{\url{https://huggingface.co/openai/whisper-large-v3}} & \textbf{14.71} & RoBERTa-XLM\footnote{\url{https://huggingface.co/cardiffnlp/twitter-xlm-roberta-base}} & \textbf{50.74 }& \textbf{48.42} & 39.10 & 46.09 \\
\bottomrule

\end{tabular}
\end{adjustbox}
\end{table}

\subsection{Results on Various Text Modality Models}
The left side of Table \ref{tab:combined_performance} compares three S2T ASR models: w2v-CTC \cite{baevski2020wav2vec}, Whisper-tiny-en \cite{radford2022whisper}, and Whisper-large-v3 \cite{radford2022whisper}. Each model was used to transcribe audio segments from the IEMOCAP dataset. Among them, Whisper-large-v3 achieved the lowest word error rate (WER) of 14.71\% and was selected as the primary S2T model for the secondary branch.

The right side of Table \ref{tab:combined_performance} compares three sentiment analysis models used to classify authentic IEMOCAP transcripts into three sentiment categories: Positive, Neutral, and Negative, where the Negative class encompasses both angry and sad emotional categories. To evaluate sentiment prediction performance, we tested three advanced Transformer-based models: DistilBERT \cite{sanh2019distilbert}, RoBERTa \cite{adoma2020comparative}, and RoBERTa-XLM \cite{conneau2019unsupervised}. Cleaned transcripts were passed through each model to generate sentiment predictions. While RoBERTa achieved the highest overall F1 score (46.12\%), RoBERTa-XLM showed the most consistent and balanced classification performance across all sentiment categories, particularly excelling in the Negative class with an F1 score of 50.74\%. Given the importance of accurately detecting negative sentiment in emotion recognition, RoBERTa-XLM was selected as the sentiment analysis model for our secondary pipeline\footnote{We also experimented with other sentiment models, such as Gemini-1.5-Flash and GPT-4o Mini, but did not observe improved performance. Since these models have not been adapted to IEMOCAP, MSP-IMPROV, or similar content-specific datasets, their performance remains suboptimal. We believe that a sentiment model better tuned to such datasets would yield improved results in the merged pipeline.}.

\subsection{Results on the Entropy-Aware Score Selection}
In this section, we first present the averaged results across different folds on the IEMOCAP dataset to compare the effectiveness of using entropy and varentropy thresholds for guiding dynamic switching between the primary and secondary pipeline models, versus not using any score selection. We then report the results for three metrics, Unweighted Accuracy (UA), Weighted Accuracy (WA), and F1 Score, for each fold on both IEMOCAP and MSP-IMPROV.

\begin{table}[t]
\centering
\caption{Average results for UA,  WA, F1 for varentropy, entropy and entropy + varentropy on the IEMOCAP dataset}
  \vspace{-3mm}
\label{tab:avg_test_ua}
\begin{tabular}{c| c |c|c|c}
\toprule
\textbf{Modality} &\textbf{Score Selection}  & \textbf{ UA (\%)} & \textbf{ WA (\%)} & \textbf{F1  (\%)}\\
\midrule
S &  w/o & 65.36 & 64.64 & 64.01 \\
S+T & Entropy  & \textbf{65.87} & 63.85 & 64.46 \\
S+T & Varentropy  & 65.56 & 64.81 & 64.24 \\
S+T & Entropy + Varentropy & 65.81 & \textbf{65.05} & \textbf{64.55} \\
\bottomrule
\end{tabular}
  \vspace{-5mm}
\end{table}

\subsubsection{Comparison With and Without Score Merging Methods}
Table~\ref{tab:avg_test_ua} lists the results on IEMOCAP with and without score selection strategies. Apparently, applying any score selection strategy, whether based on entropy, varentropy, or both of them, leads to improvements across all evaluation metrics compared to the speech-modality-only method (w/o score selection), except when using entropy alone, where the WA result (63.85\%) falls slightly short. Among the strategies, the combined \emph{Entropy + Varentropy} approach achieves the highest overall performance, with UA (65.81\%), WA (65.41\%), and F1 score (64.55\%), surpassing both individual-threshold methods and significantly outperforming the baseline. Therefore, we adopt the combined approach for its enhanced robustness and reliability for the following experiments.

\begin{table}[!htbp]
\caption{UA, WA, and F1 Score Results Across Folds For our Proposed Method on IEMOCAP Dataset}
  \vspace{-3mm}
\label{tab:iemocap_results}
\centering
\begin{adjustbox}{max width=\linewidth}
\begin{tabular}{c|ll|ll|ll}
\toprule
\textbf{Fold} 
& \multicolumn{2}{c|}{\textbf{UA (\%)}} 
& \multicolumn{2}{c|}{\textbf{WA (\%)}} 
& \multicolumn{2}{c}{\textbf{F1 Score (\%)}} \\
\cmidrule(lr){2-3} \cmidrule(lr){4-5} \cmidrule(lr){6-7}
& Before & After (Change) & Before & After (Change) & Before & After (Change) \\
\midrule
1 & 71.04 & 70.53 (\textcolor{red}{-0.51}) & 68.18 & 67.80 (\textcolor{red}{-0.38}) & 68.12 & 67.72 (\textcolor{red}{-0.40}) \\
2 & 70.86 & 70.58 (\textcolor{red}{-0.28}) & 69.30 & 69.12 (\textcolor{red}{-0.18}) & 69.91 & 69.73 (\textcolor{red}{-0.18}) \\
3 & 67.89 & 69.40 (\textcolor{green!60!black}{1.51}) & 68.81 & 70.06 (\textcolor{green!60!black}{1.25}) & 69.95 & 71.34 (\textcolor{green!60!black}{1.39}) \\
4 & 70.80 & 71.34 (\textcolor{green!60!black}{0.54}) & 67.34 & 67.71 (\textcolor{green!60!black}{0.37}) & 68.83 & 69.10 (\textcolor{green!60!black}{0.27}) \\
5 & 62.58 & 64.09 (\textcolor{green!60!black}{1.51}) & 63.03 & 64.37 (\textcolor{green!60!black}{1.34}) & 60.60 & 62.50 (\textcolor{green!60!black}{1.90}) \\
6 & 62.41 & 62.79 (\textcolor{green!60!black}{0.38}) & 62.32 & 62.96 (\textcolor{green!60!black}{0.64}) & 61.98 & 62.59 (\textcolor{green!60!black}{0.61}) \\
7 & 61.10 & 61.54 (\textcolor{green!60!black}{0.44}) & 62.12 & 62.88 (\textcolor{green!60!black}{0.76}) & 59.49 & 60.25 (\textcolor{green!60!black}{0.76}) \\
8 & 66.48 & 65.31 (\textcolor{red}{-1.17}) & 64.02 & 63.22 (\textcolor{red}{-0.80}) & 63.98 & 63.18 (\textcolor{red}{-0.80}) \\
9 & 64.43 & 64.98 (\textcolor{green!60!black}{0.55}) & 66.27 & 65.93 (\textcolor{red}{-0.32}) & 67.28 & 67.23 (\textcolor{red}{-0.05}) \\
10 & 56.02 & 57.77 (\textcolor{green!60!black}{1.55}) & 54.99 & 56.53 (\textcolor{green!60!black}{1.54}) & 50.00 & 51.95 (\textcolor{green!60!black}{1.95}) \\
\midrule
\textbf{AVG} 
& 65.36 & 65.81 (\textcolor{green!60!black}{0.45}) 
& 64.64 & 65.06 (\textcolor{green!60!black}{0.42}) 
& 64.01 & 64.56 (\textcolor{green!60!black}{0.55}) \\
\bottomrule
\end{tabular}
\end{adjustbox}
  \vspace{-3mm}
\end{table}
\begin{table}[!htbp]
\caption{UA, WA, and F1 Score Results Across Folds for our Proposed Method on MSP-IMPROV Dataset}
  \vspace{-3mm}
\label{tab:msp_results}
\centering
\begin{adjustbox}{max width=\linewidth}
\begin{tabular}{c|ll|ll|ll}
\toprule
\textbf{Fold} 
& \multicolumn{2}{c|}{\textbf{UA (\%)}} 
& \multicolumn{2}{c|}{\textbf{WA (\%)}} 
& \multicolumn{2}{c}{\textbf{F1 Score (\%)}} \\
\cmidrule(lr){2-3} \cmidrule(lr){4-5} \cmidrule(lr){6-7}
& Before & After (Change) & Before & After (Change) & Before & After (Change) \\
\midrule
1 & 58.80 & 58.41 (\textcolor{red}{-0.39}) & 65.02 & 64.55 (\textcolor{red}{-0.47}) & 58.55 & 58.08 (\textcolor{red}{-0.47}) \\
2 & 50.74 & 52.88 (\textcolor{green!60!black}{2.14}) & 59.59 & 61.07 (\textcolor{green!60!black}{1.38}) & 51.81 & 54.30 (\textcolor{green!60!black}{2.49}) \\
3 & 50.16 & 52.05 (\textcolor{green!60!black}{1.89}) & 65.33 & 65.9 (\textcolor{green!60!black}{0.57}) & 50.55 & 52.73 (\textcolor{green!60!black}{2.18}) \\
4 & 49.76 & 50.95 (\textcolor{green!60!black}{1.19}) & 54.69 & 55.84 (\textcolor{green!60!black}{1.15}) & 51.12 & 52.5 (\textcolor{green!60!black}{1.38}) \\
5 & 59.17 & 59.06 (\textcolor{red}{-0.11}) & 63.1 & 63.1 (0.00) & 60.72 & 60.48 (\textcolor{red}{-0.24}) \\
6 & 44.71 & 46.11 (\textcolor{green!60!black}{1.40}) & 50.17 & 51.16 (\textcolor{green!60!black}{0.99}) & 43.99 & 45.7 (\textcolor{green!60!black}{1.71}) \\
\midrule
\textbf{AVG} 
& 52.22 & 53.24 (\textcolor{green!60!black}{1.02}) 
& 59.67 & 60.27 (\textcolor{green!60!black}{0.69}) 
& 52.79 & 53.97 (\textcolor{green!60!black}{1.18}) \\
\bottomrule
\end{tabular}
\end{adjustbox}
  \vspace{-3mm}
\end{table}

\subsubsection{IEMOCAP Dataset Results} Table~\ref{tab:iemocap_results} shows fold-wise performance of the proposed score selection method on IEMOCAP. Results are compared before and after applying the entropy + varentropy-based score selection. 
Seven out of ten folds (Folds 3–7, 9, and 10) exhibit consistent improvements across all three evaluation metrics, demonstrating that the merge strategy, through confidence-driven switching, leveraged textual sentiment cues to correct misclassifications. Even in folds like Fold 6 with strong baseline accuracy, the algorithm yielded marginal improvements or stable results, demonstrating non-destructive behavior.

\subsubsection{MSP-IMPROV Dataset Results} To further validate the generalizability of our approach, we applied the proposed score selection method to the MSP-IMPROV dataset. The speech scores were obtained from the wav2vec2 model fine-tuned on the MSP-IMPROV dataset, while the text scores were obtained using the same text model as used for IEMOCAP.
Results from Table~\ref{tab:msp_results} similarly show improvements. Folds 2-4, and 6 in particular saw significant uplifts, suggesting that the entropy + varentropy-based merge is not overfit to IEMOCAP's structure and retains robustness under distributional shift. The merge mechanism was able to dynamically defer to the secondary sentiment pipeline when confidence was low, preserving precision without introducing instability.

The final assessment shows consistent performance gains using our entropy-aware score selection strategy across both datasets. Although some folds experienced slight performance drops, this may be due to low-confidence predictions or threshold misalignment in certain emotion classes. On average, improvements in F1 score range from 0.5\% to 1.2\%, indicating the effectiveness of the merging framework while highlighting room for further enhancement. Future work could explore more powerful text-based models to strengthen the utility of the secondary sentiment signal. Additionally, we observe that entropy and varentropy thresholds differ across emotion classes, suggesting that class-specific or dynamically adaptive thresholding could further refine fusion decisions.

\begin{figure}[t]
  \centering
  \includegraphics[width=1\linewidth]{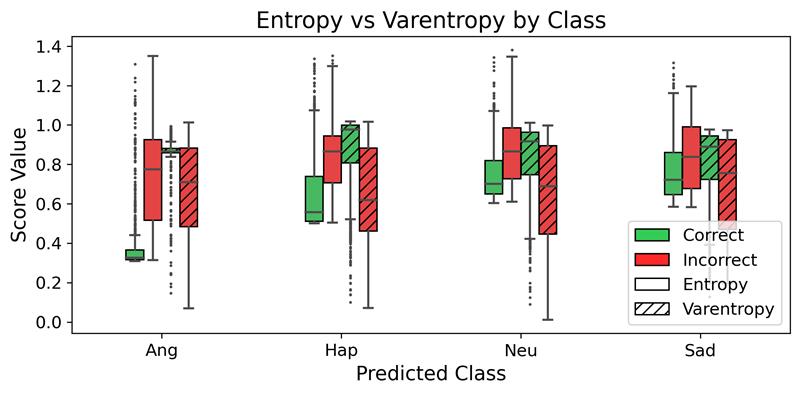}
  \caption{Distributions of entropy (solid color-filled bars) and varentropy (hatched color-filled bars) for each predicted emotion class on Fold 3 of the IEMOCAP dataset. For each class, values are shown separately for correctly predicted (green) and incorrectly predicted (red) samples.}
  \label{fig:fold_2}
  \vspace{-4mm}
\end{figure}

\subsubsection{Further Analysis}
As explained in Section \ref{sec:entropy-aware}, lower entropy and higher varentropy are indicative of more accurate predictions, and the search space of these metrics is dependent on the emotion class. Therefore, it is insightful to visualize how their distributions vary across classes. Figure \ref{fig:fold_2} plots the entropy (solid color-filled bars) and varentropy (hatched color-filled bars) for each predicted class on Fold 3 of the IEMOCAP dataset. For each class, the values are shown separately for correctly predicted (green) and incorrectly predicted (red) samples.
It is obvious that lower entropy and higher varentropy correlate with more accurate predictions. For entropy, the green bars (correct predictions) are consistently lower than the red bars (incorrect predictions) across all four classes, suggesting that lower entropy values are associated with greater confidence and correctness. In the case of varentropy, the green hatched bars (correct predictions) are generally higher or more centrally distributed than the red hatched bars (incorrect predictions), indicating that higher varentropy values likewise correspond to greater prediction confidence and accuracy.

Moreover, each emotion class demonstrates a distinct range of entropy and varentropy values, reinforcing the need for a per-class thresholding strategy. A single global threshold would fail to capture these class-specific patterns effectively. Additionally, the merging process can occasionally turn a correct prediction into an incorrect one. Therefore, choosing an appropriate threshold becomes a trade-off: it must balance maximizing the detection of incorrect predictions while minimizing the erroneous rejection of correct ones. This trade-off motivates the use of accuracy as the evaluation metric during parameter selection.

\section*{Conclusion}
In this study, we proposed a multimodal score selection methodology that combines a primary wav2vec2-based speech branch with a secondary Whisper + RoBERTa-XLM sentiment branch for SER. For score selection between the two branches, we leveraged entropy and varentropy thresholds to identify uncertain predictions and dynamically switched to the secondary pipeline for improved reliability. Experiments on both the IEMOCAP and MSP-IMPROV datasets demonstrated clear improvements in accuracy, robustness, and stability across emotional classes, particularly in challenging cases involving conflicting emotional cues. Overall, the proposed late fusion strategy offers a computationally efficient, flexible, and reliable pipeline, outperforming single-modality systems and fixed fusion strategies.

\bibliographystyle{IEEEtran}
\bibliography{mybib}

\begin{thebibliography}{10}
\providecommand{\url}[1]{#1}
\csname url@samestyle\endcsname
\providecommand{\newblock}{\relax}
\providecommand{\bibinfo}[2]{#2}
\providecommand{\BIBentrySTDinterwordspacing}{\spaceskip=0pt\relax}
\providecommand{\BIBentryALTinterwordstretchfactor}{4}
\providecommand{\BIBentryALTinterwordspacing}{\spaceskip=\fontdimen2\font plus
\BIBentryALTinterwordstretchfactor\fontdimen3\font minus
  \fontdimen4\font\relax}
\providecommand{\BIBforeignlanguage}[2]{{%
\expandafter\ifx\csname l@#1\endcsname\relax
\typeout{** WARNING: IEEEtran.bst: No hyphenation pattern has been}%
\typeout{** loaded for the language `#1'. Using the pattern for}%
\typeout{** the default language instead.}%
\else
\language=\csname l@#1\endcsname
\fi
#2}}
\providecommand{\BIBdecl}{\relax}
\BIBdecl

\bibitem{busso2008iemocap}
C.~Busso, M.~Bulut, C.-C. Lee, A.~Kazemzadeh, E.~Mower, S.~Kim, J.~N. Chang,
  S.~Lee, and S.~S. Narayanan, ``Iemocap: Interactive emotional dyadic motion
  capture database,'' \emph{Language resources and evaluation}, vol.~42, pp.
  335--359, 2008.

\bibitem{maji2023multimodal}
B.~Maji, M.~Swain, R.~Guha, and A.~Routray, ``Multimodal emotion recognition
  based on deep temporal features using cross-modal transformer and
  self-attention,'' in \emph{ICASSP 2023-2023 IEEE International Conference on
  Acoustics, Speech and Signal Processing (ICASSP)}.\hskip 1em plus 0.5em minus
  0.4em\relax IEEE, 2023, pp. 1--5.

\bibitem{bertero2017first}
D.~Bertero and P.~Fung, ``A first look into a convolutional neural network for
  speech emotion detection,'' in \emph{2017 IEEE international conference on
  acoustics, speech and signal processing (ICASSP)}.\hskip 1em plus 0.5em minus
  0.4em\relax IEEE, 2017, pp. 5115--5119.

\bibitem{khalil2019speech}
R.~A. Khalil, E.~Jones, M.~I. Babar, T.~Jan, M.~H. Zafar, and T.~Alhussain,
  ``Speech emotion recognition using deep learning techniques: A review,''
  \emph{IEEE access}, vol.~7, pp. 117\,327--117\,345, 2019.

\bibitem{baevski2020wav2vec}
A.~Baevski, Y.~Zhou, A.~Mohamed, and M.~Auli, ``wav2vec 2.0: A framework for
  self-supervised learning of speech representations,'' \emph{Advances in
  neural information processing systems}, vol.~33, pp. 12\,449--12\,460, 2020.

\bibitem{hsu2021hubert}
W.-N. Hsu, Y.-H.~H. Tsai, B.~Bolte, R.~Salakhutdinov, and A.~Mohamed, ``Hubert:
  How much can a bad teacher benefit asr pre-training?'' in \emph{ICASSP
  2021-2021 IEEE International Conference on Acoustics, Speech and Signal
  Processing (ICASSP)}.\hskip 1em plus 0.5em minus 0.4em\relax IEEE, 2021, pp.
  6533--6537.

\bibitem{chen2022wavlm}
S.~Chen, C.~Wang, Z.~Chen, Y.~Wu, S.~Liu, Z.~Chen, J.~Li, N.~Kanda,
  T.~Yoshioka, X.~Xiao \emph{et~al.}, ``Wavlm: Large-scale self-supervised
  pre-training for full stack speech processing,'' \emph{IEEE Journal of
  Selected Topics in Signal Processing}, vol.~16, no.~6, pp. 1505--1518, 2022.

\bibitem{poria2019emotion}
S.~Poria, N.~Majumder, R.~Mihalcea, and E.~Hovy, ``Emotion recognition in
  conversation: Research challenges, datasets, and recent advances,''
  \emph{IEEE access}, vol.~7, pp. 100\,943--100\,953, 2019.

\bibitem{Baltruvsaitis2018Multimodal}
T.~Baltru{\v{s}}aitis, C.~Ahuja, and L.-P. Morency, ``Multimodal machine
  learning: A survey and taxonomy,'' \emph{IEEE transactions on pattern
  analysis and machine intelligence}, vol.~41, no.~2, pp. 423--443, 2018.

\bibitem{poria2018multimodal}
S.~Poria, N.~Majumder, D.~Hazarika, E.~Cambria, A.~Gelbukh, and A.~Hussain,
  ``Multimodal sentiment analysis: Addressing key issues and setting up the
  baselines,'' \emph{IEEE Intelligent Systems}, vol.~33, no.~6, pp. 17--25,
  2018.

\bibitem{zadeh2017tensor}
A.~Zadeh, M.~Chen, S.~Poria, E.~Cambria, and L.-P. Morency, ``Tensor fusion
  network for multimodal sentiment analysis,'' \emph{arXiv preprint
  arXiv:1707.07250}, 2017.

\bibitem{lu2020speech}
Z.~Lu, L.~Cao, Y.~Zhang, C.-C. Chiu, and J.~Fan, ``Speech sentiment analysis
  via pre-trained features from end-to-end asr models,'' in \emph{ICASSP
  2020-2020 IEEE International Conference on Acoustics, Speech and Signal
  Processing (ICASSP)}.\hskip 1em plus 0.5em minus 0.4em\relax IEEE, 2020, pp.
  7149--7153.

\bibitem{devlin2019bert}
J.~Devlin, M.-W. Chang, K.~Lee, and K.~Toutanova, ``Bert: Pre-training of deep
  bidirectional transformers for language understanding,'' in \emph{Proceedings
  of the 2019 conference of the North American chapter of the association for
  computational linguistics: human language technologies, volume 1 (long and
  short papers)}, 2019, pp. 4171--4186.

\bibitem{liu2019roberta}
Y.~Liu, ``Roberta: A robustly optimized bert pretraining approach,''
  \emph{arXiv preprint arXiv:1907.11692}, vol. 364, 2019.

\bibitem{katsaggelos2015audiovisual}
A.~K. Katsaggelos, S.~Bahaadini, and R.~Molina, ``Audiovisual fusion:
  Challenges and new approaches,'' \emph{Proceedings of the IEEE}, vol. 103,
  no.~9, pp. 1635--1653, 2015.

\bibitem{chen2024modality}
C.~Chen and P.~Zhang, ``Modality-collaborative transformer with hybrid feature
  reconstruction for robust emotion recognition,'' \emph{ACM Transactions on
  Multimedia Computing, Communications and Applications}, vol.~20, no.~5, pp.
  1--23, 2024.

\bibitem{georgescu2024exploring}
A.-L. Georgescu, G.-I. Chivu, and H.~Cucu, ``Exploring fusion techniques for
  multimodal emotion recognition,'' in \emph{2024 15th International Conference
  on Communications (COMM)}.\hskip 1em plus 0.5em minus 0.4em\relax IEEE, 2024,
  pp. 1--6.

\bibitem{song2018decision}
K.-S. Song, Y.-H. Nho, J.-H. Seo, and D.-s. Kwon, ``Decision-level fusion
  method for emotion recognition using multimodal emotion recognition
  information,'' in \emph{2018 15th international conference on ubiquitous
  robots (UR)}.\hskip 1em plus 0.5em minus 0.4em\relax IEEE, 2018, pp.
  472--476.

\bibitem{busso2016msp}
C.~Busso, S.~Parthasarathy, A.~Burmania, M.~AbdelWahab, N.~Sadoughi, and E.~M.
  Provost, ``Msp-improv: An acted corpus of dyadic interactions to study
  emotion perception,'' \emph{IEEE Transactions on Affective Computing},
  vol.~8, no.~1, pp. 67--80, 2016.

\bibitem{camacho2017role}
J.~Camacho-Collados and M.~T. Pilehvar, ``On the role of text preprocessing in
  neural network architectures: An evaluation study on text categorization and
  sentiment analysis,'' \emph{arXiv preprint arXiv:1707.01780}, 2017.

\bibitem{radford2023robust}
A.~Radford, J.~W. Kim, T.~Xu, G.~Brockman, C.~McLeavey, and I.~Sutskever,
  ``Robust speech recognition via large-scale weak supervision,'' in
  \emph{International conference on machine learning}.\hskip 1em plus 0.5em
  minus 0.4em\relax PMLR, 2023, pp. 28\,492--28\,518.

\bibitem{barbieri2021xlm}
F.~Barbieri, L.~E. Anke, and J.~Camacho-Collados, ``Xlm-t: Multilingual
  language models in twitter for sentiment analysis and beyond,'' \emph{arXiv
  preprint arXiv:2104.12250}, 2021.

\bibitem{radford2022whisper}
\BIBentryALTinterwordspacing
A.~Radford, J.~W. Kim, T.~Xu, G.~Brockman, C.~McLeavey, and I.~Sutskever,
  ``Robust speech recognition via large-scale weak supervision,'' 2022.
  [Online]. Available: \url{https://arxiv.org/abs/2212.04356}
\BIBentrySTDinterwordspacing

\bibitem{sanh2019distilbert}
V.~Sanh, L.~Debut, J.~Chaumond, and T.~Wolf, ``Distilbert, a distilled version
  of bert: smaller, faster, cheaper and lighter,'' \emph{arXiv preprint
  arXiv:1910.01108}, 2019.

\bibitem{adoma2020comparative}
A.~F. Adoma, N.-M. Henry, and W.~Chen, ``Comparative analyses of bert, roberta,
  distilbert, and xlnet for text-based emotion recognition,'' in \emph{2020
  17th international computer conference on wavelet active media technology and
  information processing (ICCWAMTIP)}.\hskip 1em plus 0.5em minus 0.4em\relax
  IEEE, 2020, pp. 117--121.

\bibitem{conneau2019unsupervised}
A.~Conneau, K.~Khandelwal, N.~Goyal, V.~Chaudhary, G.~Wenzek, F.~Guzm{\'a}n,
  E.~Grave, M.~Ott, L.~Zettlemoyer, and V.~Stoyanov, ``Unsupervised
  cross-lingual representation learning at scale,'' \emph{arXiv preprint
  arXiv:1911.02116}, 2019.

\end{thebibliography}

\end{document}